\documentclass[12pt]{article}
\usepackage[utf8]{inputenc}
\usepackage[T1]{fontenc}
\usepackage{amsmath}
\usepackage{graphicx}
\usepackage{txfonts}
\usepackage{color}
\usepackage[version=4]{mhchem}
\usepackage{authblk}
\usepackage{caption}
\usepackage{siunitx}

\makeatletter
\renewcommand{\@maketitle}{%
  \vspace*{-80pt}
  {\parindent0pt \Large \bfseries \@title \par}
  \vspace{5pt}
  {\parindent0pt \normalsize
  \linespread{1.5}\selectfont 
  \@author \par}
  \vspace{5pt}
  {\footnotesize
  \@date \par} 
}
\renewcommand{\section}{\@startsection{section}{1}{0pt}%
  {-12pt}
  {6pt}
  {\normalfont\normalsize\bfseries}} 
\makeatother

\newenvironment{sciabstract}{%
\begin{quote} \bf}
{\end{quote}}

\newcommand{\spacing}[1]{\renewcommand{\baselinestretch}{#1}\large\normalsize}
\spacing{2}

\title{Optical multistability in a compact microcavity enabled by near-exceptional coupling}

\author[1]{Zhen Liu}
\author[1]{Xuefan Yin}
\author[2,3]{Andrey Bogdanov}
\author[1]{Yujia Nie}
\author[1]{Yi Zuo}
\author[1,4]{Hongbin Li}
\author[1,*]{Feifan Wang}
\author[1,4,*]{Chao Peng}
\affil[1]{State Key Laboratory of Advanced Optical Communication Systems and Networks, School of Electronics \& Frontiers Science Center for Nano-optoelectronics, Peking University, Beijing 100871, China}
\affil[2]{Qingdao Innovation and Development Center, Harbin Engineering University, Qingdao 266000, Shandong, China}
\affil[3]{School of Physics and Engineering, ITMO University, Saint Petersburg 197101, Russia}
\affil[4]{Peng Cheng Laboratory, Shenzhen 518055, China}
\affil[*]{To whom correspondence should be addressed; E-mail:    Feifan Wang (wangfeifan@pku.edu.cn), Chao Peng (pengchao@pku.edu.cn).}

\date{}

\begin{document}

\baselineskip 24pt
\maketitle

\newpage
\begin{sciabstract}
Multistability --- the emergence of multiple stable states under identical conditions --- is a hallmark of nonlinear complexity and an enabling mechanism for multilevel optical memory and photonic computing. Its realization in a compact footprint, however, is limited by intrinsically weak optical nonlinearities and the enlarged free spectral range that raises the multistability threshold. Here, we overcome this constraint by engineering a pair of spectrally close, ultra-high-$Q$ resonances in a photonic crystal microcavity. Leveraging structural perturbations that deliberately introduce non-Hermitian coupling through a shared radiation channel, we drive the resonances toward an exceptional point with nearly degenerate wavelengths and balanced quality factors approaching $10^6$. This configuration substantially enhances thermo-optical nonlinearity and produces pronounced tristability and hysteresis loops within a footprint of \qty{20}{\micro m} at input powers below \qty{240}{\micro\watt}. We further demonstrate proof-of-concept optical random-access memory through controlled switching among multistable states. These results establish a general strategy for nonlinear microcavities to achieve energy-efficient multistability for reconfigurable all-optical memories, logic, and neuromorphic processors.

\end{sciabstract}

\section*{Introduction}

Multistability is a hallmark of complex nonlinear interactions ubiquitously observed across biological \cite{leopold1999MultistablePhenomenaChanging,ferrell1998biochemical,scheffer2001catastrophic,rietkerk_evasion_2021}, mechanical \cite{ravelet2004multistability,meeussen_multistable_2023}, and electrical \cite{arecchi1982hopping,hahnloser_digital_2000} systems, and is also considered essential physics for artificial intelligence and neuromorphic computing \cite{chua1988cellular,foss1996multistability,angeli2004detection,ferrell2002self,hancock_metastability_2025}. As a reduced form involving two stable optical states, optical bistability has been successfully realized with a low threshold, fast response, and compactness for large-scale integration, enabling applications from all-optical switching, memory, to photonic neural networks \cite{goldstone_intrinsic_1984,soljavcic2002optimal,hill_fast_2004,nozaki_ultralow-power_2012,nishida2023OpticalBistabilityNanosilicon,lampande_positive-feedback_2024,skripka_intrinsic_2025}. With almost all advantages of bistability preserved, optical multistability further offers access to multi-bit stable states and thus paves a promising pathway towards advanced photonic computing with complexity and flexibility. However, realizing this behavior generally requires a multi-level system with strong inter-level interactions, which are inherently weak in typical optical platforms. As a result, strategies including large device footprints to reduce free-spectral range, heterogeneous materials to promote nonlinearity, cascade configurations or stringent operating conditions \cite{kitano1981optical,cecchi1982observation,paraiso_multistability_2010,keijsers2025continuous} are employed to enhance the inter-level interactions, leaving a compact, efficient, and elegant chip-integrated approach to optical multistability still out of reach.

A promising route towards on-chip optical multistability is to exploit a set of nearly degenerate resonances within a compact microcavity, where inter-resonance interactions are naturally enhanced. In photonic crystal (PhC) microcavities operating within the light cone, structural symmetry ensure degenerate cavity modes \cite{joannopoulos1997photonic,fan2002AnalysisGuidedResonances,doiron2022RealizingSymmetryguaranteedPairs,chen2025ObservationChiralEmission}, while controlled asymmetry can couple them and open up radiation channels for efficient excitation and emission \cite{nguyen2018symmetry,hamel2015spontaneous,yin2023topological,koshelev2018AsymmetricMetasurfacesHighQ,wang2023BrillouinZoneFolding}. Generally, achieving a strong nonlinear response requires ultra-high quality factors ($Q$) \cite{purcell1946SpontaneousEmissionProbabilities,akahane2003high,soljavcic2004enhancement,koshelev2020subwavelength}, yet high-$Q$ modes are intrinsically difficult to access without carefully engineering their radiation to satisfy critical coupling \cite{haus1984waves,cai2000observation,suh2004TemporalCoupledmodeTheory,chong2010CoherentPerfectAbsorbers,soleymani2022ChiralDegeneratePerfect,hinney2024efficient,schiattarella2024directive}. Note that when multiple modes couple to a common radiation channel, coherent and dissipative interactions can drive the system toward an exceptional point (EP) where eigenmodes coalesce, producing hybridized modes with slight frequency splitting and nearly identical while tunable linewidths \cite{moiseyev2011NonHermitianQuantumMechanics,heiss2012PhysicsExceptionalPoints,zhen2015SpawningRingsExceptional,miri2019ExceptionalPointsOptics,el-ganainy2018NonHermitianPhysicsPT,zhang2025NonHermitianSingularitiesScattering}. In this near-EP regime, multiple modes can be efficiently excited while maintaining a stable inter-mode coupling --- the key to achieving robust optical multistability that operates free from modulation instability and chaos.

Here, we realize optical multistability in a silicon PhC cavity with compact footprint by efficiently exciting a pair of symmetry-originated, near-degenerate optical modes, engineered to exhibit an appropriate spectral splitting together with high yet balanced $Q$ factors --- a regime we refer to as near-exceptional coupling (NEC). Specifically, structural perturbations introduce both non-radiative and radiative coupling between two cavity modes, driving them toward an EP with closely spaced wavelengths and nearly identical excitation efficiencies. Leveraging the ultra-high-$Q$ nature ($\sim10^6$) and efficient excitation achieved in the NEC regime, we demonstrate distinct thermo-optic multistability within a cavity core whose footprint is smaller than \qty{20}{\micro\meter}, at input powers as low as \qty{240}{\micro\watt}. We further present a proof-of-concept optical random-access memory through time-domain switching between the stable states. This approach provides a compact and energy-efficient building block for optical multistability, paving the way toward reconfigurable photonic logic, all-optical signal processing, and memory devices.

\section*{Principle and design}

Our system is schematically illustrated in Fig.~1a. The cavity is patterned on a silicon-on-insulator (SOI) wafer and comprises a hexagonal PhC core surrounded by a gradually stretched lattice that forms an in-plane photonic bandgap for optical confinement (see \textcolor{black}{Methods} for details). Two nearly degenerate modes, Res A and Res B(left inset, Fig. 1a), are laterally confined within the core region (red shaded) and couple to the radiation continuum, enabling efficient out-of-plane excitation and emission (red arrow). In such a system, absorption-induced heating shifts the mode wavelengths, providing the nonlinear feedback required for multistable dynamics (right inset, Fig.~1a, see \textcolor{black}{Supplementary Sections 1 and 2} for details). As the resonant characteristics govern the nonlinear response, we engineer the cavity modes to be closely spaced yet spectrally resolvable with balanced radiative decay rates, supporting strong intracavity fields and enhanced thermo-optical nonlinearity for effective multi-level state transitions.

We start from a pair of degenerate modes with frequencies of $\omega_0$. A Hermitian perturbation of strength $\delta$ splits the frequencies slightly, while the modes couple to the radiation bath with strengths $\gamma_{1,2}$. The system is then described by the Hamiltonian $\mathbf{H}_0$ perturbed by $\Delta \mathbf{H}_\mathrm{split}$ and $\Delta \mathbf{H}_\mathrm{rad}$ as:
\vspace{0.5cm}
\begin{align} \mathbf{H}_0+\Delta \mathbf{H}_\mathrm{split}+\Delta \mathbf{H}_\mathrm{rad}=\begin{pmatrix}\omega_0 & 0\\ 0& \omega_0\end{pmatrix}+\begin{pmatrix}\delta & 0\\ 0& -\delta\end{pmatrix}+i\begin{pmatrix}\gamma_1 & \gamma_{12} \\ \gamma_{21} & \gamma_2\end{pmatrix} \nonumber
\end{align}

Here the non-radiative losses are excluded for simplicity. By deliberately selecting the ways of symmetry-breaking, both Res A and B radiate into the same linear polarization, ensuring $\gamma_{12} = \gamma_{21} = \sqrt{\gamma_1 \gamma_2}$ according to the law of energy conservation \cite{suh2004TemporalCoupledmodeTheory}. Further assuming the cavity modes are coupled with equal rates as $\gamma_{1,2}=\gamma$, we find the system supports two states with eigenfrequencies as:
\begin{equation*}
\omega_{\pm}=\omega_0+i\gamma \pm \sqrt{\delta^2-\gamma^2}
\end{equation*}
The imaginary parts represent the decay rates of the eigenstates through radiation. The system reaches the EP when $\gamma=\delta$, corresponding to a non-Hermitian degeneracy where the two eigenstates fully coalesce \cite{heiss2012PhysicsExceptionalPoints,miri2019ExceptionalPointsOptics,el-ganainy2018NonHermitianPhysicsPT}. For a fixed Hermitian perturbation strength $\delta$, increasing radiation rates as $\gamma=0 \to \delta$ would drive the system to approach the EP gradually. In this transition, the frequency splitting decreases, while the radiation rates of the two modes slightly increase but remain nearly identical. We refer to this regime as NEC.

The two-fold degeneracy arises from folding of the Brillouin zone in a simple hexagonal lattice (Fig.~1c) \cite{wang2023BrillouinZoneFolding}. Specifically, transverse-electric (TE) bands in the unit cell (purple curves) extend from the Brillouin-zone center to distinct $K$ points ($K$, $K^\prime$), which fold to the $\Gamma$ point in the supercell perspective and coalesce (indicated by arrow in Fig.~1c, see \textcolor{black}{Supplementary Section 3} for details). In a finite-sized structure, the bands are discretized into two sets of cavity modes, including Res A and Res B. The degeneracy is naturally lifted due to the breaking of infinite periodicity, while the modes remain non-radiative.  We then apply selective structural perturbations to control the frequency splitting and radiative coupling. The coupling of Res A and B to the radiation continuum (Fig.~1d) can be independently introduced by shifting air holes along the $y$ direction: moving the central hole in each supercell ($d_1$ perturbation, Fig.~1e) only couples Res A to $x$-polarized radiation, while staggered shifts of the other two sets of holes ($d_2$ perturbation, Fig.~1e) generate the same polarized radiation but merely to Res B. A detailed analysis of these perturbations is provided in \textcolor{black}{Supplementary Sections~4 and 5}. Together, these manipulations mediate the mode splitting and balance the radiation rates, thus fulfilling the requirements for NEC.

We evaluate the eigenstate evolution using an effective Hamiltonian that incorporates the influences of perturbations $d_1$ and $d_2$ (Fig.~1f, see \textcolor{black}{Supplementary Sections~4 and 5} for details). We define a normalized coupling parameter $\gamma_\mathrm{norm}$ to quantify the radiative coupling strength, which is proportional to $d_1^2+d_2^2$ and reaches 1 at the EP. Without perturbation, Res A and B exhibit a small initial mode splitting due to finite-size effects, while remaining non-radiative (yellow-shaded region). Shifting the air holes renders them radiative (gray-shaded region), driving the system toward the EP. For small $\gamma_\mathrm{norm}$, the radiative $Q$ remains too high for efficient excitation. By increasing $d_1$ and $d_2$,  a larger $\gamma_\mathrm{norm}$ enhances the intracavity energy, which reaches a maximum under the critical coupling condition \cite{haus1984waves,cai2000observation}, where radiative and non-radiative losses are balanced ($\mathrm{Im}(\omega_\pm) = \gamma_\mathrm{non\text{-}rad}$). Here, $\gamma_\mathrm{non\text{-}rad}$ accounts for unavoidable scattering and absorption \cite{iadanza2020ModelThermoopticNonlinear,dinu2003ThirdorderNonlinearitiesSilicon}. In the NEC regime, $\mathrm{Im}(\omega_\pm) \sim \gamma_\mathrm{ non\text{-}rad}$ allows both modes to be efficiently and equally excited, pronouncedly enhancing optical nonlinearity. Fig.~1g shows the calculated cavity energy under excitation by using temporal coupled-mode theory (TCMT) (see \textcolor{black}{Supplementary Section~1}). As $\gamma_\mathrm{norm}$ increases, the resonant peaks grow while their splitting narrows (left three panels), favorable for promoting the nonlinear effect. Beyond this, the peaks coalesce, and the multi-level feature vanishes (last panel). The NEC regime thus corresponds to a balance between mode splitting and radiation, optimal for distinct optical multistability.
The eigenstate evolution is numerically validated (see \textcolor{black}{Methods}) by applying $d_1$ and $d_2$, following a Bernoulli lemniscate in Fig.~1h on which either real or imaginary parts of eigenfrequencies coalesce. The results show excellent agreement with theory predictions (Fig.~1i). More details are provided in \textcolor{black}{Supplementary Section~5}.

We further elaborate on the thermo-optic effects in our system \cite{iadanza2020ModelThermoopticNonlinear,carmon2004DynamicalThermalBehavior,gao2022ProbingMaterialAbsorption,barulin2024ThermoOpticalBistabilityEnabled}. When the cavity is excited by a laser, the absorbed energy heats the cavity, rendering the refractive index intensity-dependent (left panel, Fig.~2a). The heating mainly originates from linear absorption due to dangling bands and surface roughness \cite{iadanza2020ModelThermoopticNonlinear,gao2022ProbingMaterialAbsorption}, as well as two-photon absorption and free-carrier absorption \cite{dinu2003ThirdorderNonlinearitiesSilicon, uesugi2006InvestigationOpticalNonlinearities}.  Stronger resonant energy enhances the absorption, leading to a larger thermal shift. Three steady states (right panel, Fig.~2a) can coexist under the same excitation and power and wavelength (dashed line): the ``cold state" (A) corresponding to an off-resonant condition with negligible absorption; the ``warm state" (B) and ``hot state" (C) where strong absorption (red shaded regions in peaks) induces significant thermo-optic shifts(red arrows) that sustain resonance at the working wavelength. Transitions among these three steady states can be triggered by tuning the input power or wavelength following specific trajectories.

Thermo-optical nonlinearity gives rise to multistability with characteristic hysteresis as the input power varies \cite{carmon2004DynamicalThermalBehavior,bulgakov2019NonlinearResponseOptical,cotrufo2024PassiveBiasfreeNonreciprocal,shadrina2025ThermoopticBistabilityTwodimensional,marisova2025multistable_theory}. In theory, the system reduces to a single-variable model described by temperature, with equilibrium defined by the vanishing of its time derivative. The equilibrium curve separates the parameter-variable space into heating (red) and cooling (blue) regions, illustrated in Fig.~2b. Without optical modes, the curve is monotonic, yielding only one stable state for each input power. In contrast, our system hosts two near-degenerate modes, which enhance absorption and deform the response curve into a multivalued form. At a given input power (gray line), five equilibria appear, among which three (A–C) are stable under temperature perturbation, corresponding to cold, warm, and hot states in Fig.~2a. Under adiabatic parameter variation, the system follows stable branches and undergoes abrupt transitions at bifurcation points(arrows on the curve), thus producing two hysteresis loops as a signature of optical multistability.  More details are presented in \textcolor{black}{Supplementary Sections~1 and 2}.

To quantify the role of the NEC region in promoting multistability, we calculate phase diagrams and hysteresis curves using the TCMT model with thermo-optical nonlinearity (Figs.~2c-d; see \textcolor{black}{Supplementary Sections~1 and 2}). For simplicity, critical coupling is assumed at the EP, i.e., $\mathrm{Im}(\omega_{EP}) = \gamma_\mathrm{ non\text{-}rad}$, to maximize field enhancement \cite{soleymani2022ChiralDegeneratePerfect, schiattarella2024directive}. The evolution of the two resonant peaks produces distinct phase diagrams. At small $\gamma_\mathrm{norm}$ (first panels in Fig. 2c, $\gamma_\mathrm{norm} = 0.1$), each peak shows a bistable horn-like region. As $\gamma_\mathrm{norm}$ increases to 0.4 (second panel), these regions overlap, forming a multistable zone with three coexisting stable states. The bistability threshold decreases as resonant energy grows. In the NEC regime ($\gamma_\mathrm{norm} = 0.8$, third panel), the threshold drops further due to reduced frequency splitting. As $\gamma_\mathrm{norm}$ approaches unity ($0.95$, last panel), the peaks merge and the multistable region narrows, marking the disappearance of distinct multi-level operation, highlighting the optimal NEC window for pronounced optical multistability. These phenomena are also confirmed by the power hysteresis curves shown in the bottom panels in Fig.~2d.

Summarizing these trends, as $\gamma_\mathrm{norm}$ increases from 0 (non-radiative) to 1 (EP), the multistability threshold power continuously decreases (red curve, Fig.~2e), while the width of the multistable region initially broadens, peaks near $\gamma_\mathrm{norm} \approx 0.8$, and then narrows as the system approaches the EP (blue curve, Fig.~2e). This behavior reflects the evolution of the linear spectra in Fig.~1g: increasing $\gamma_\mathrm{norm}$ drives the system from the under-coupled regime toward critical coupling, enhancing on-resonance field buildup, while the two resonant peaks converge into a single square-Lorentzian lineshape at the EP. The NEC region, therefore, supports multistable hysteresis at low threshold power, whereas a further approach to the EP reduces multistability, ultimately leaving the system in a simple bistable regime.

\section*{Sample fabrication and experimental setup}

To realize multistability experimentally, we fabricate PhC cavities on a silicon-on-insulator (SOI) wafer with a 220-nm top silicon layer, following the aforementioned design principles. A representative top-view SEM image is shown in Fig.~3a. The cavity incorporates a three-layer heterostructure to enhance absorptive heating, with thermo-optic effects predominantly arising in the core region (red), which localizes the majority of the resonant energy. This region is defined by a lattice constant before perturbation of $a=400$ nm. At the boundary (blue), the lattice is stretched by 12.5\% in the radial direction to form a photonic bandgap (PbG) region, while a gradient-stretching region (yellow) provides an adiabatic transition that suppresses scattering losses \cite{song2005UltrahighQPhotonicDoubleheterostructurea}  (see \textcolor{black}{Methods}). Cavities were patterned by electron-beam lithography and etched via inductively coupled plasma. To enhance optical confinement and reduce thermal dissipation, the buried oxide layer was removed by immersing the chip in 49\% hydrofluoric acid, thereby forming a suspended membrane cavity. A zoom-in SEM of the core (Fig.~3a, inset) confirms the smooth, well-defined edges of the air holes.
\newline

NEC is achieved by controlled perturbations $d_1$ and $d_2$ in the cavity. The key control lies in tuning them to balance radiative and absorptive losses. As the perturbations require nanometer-scale precision, the fabricated samples are inevitably biased toward but not exactly to the design values. To compensate, we scan two structural degrees of freedom at the finest steps during fabrication to enable the most effective excitation of both modes (see \textcolor{black}{Methods} for more details).

We use a thermally stabilized confocal measurement system for characterization (Fig.~3b). 
The system comprises a commercial tunable C+L-band continuous-wave laser, focused onto the sample via an objective. The reflected signal is magnified by a 4$f$ system and detected by a photodetector. Two linear polarizers (LP1, LP2) provide cross-polarization filtering, strongly suppressing background reflections while preserving cavity radiation \cite{galli2009light}. The same setup is used for multistability measurements: by scanning the excitation laser power up and down at different wavelengths, we record the hysteretic response of the cavity's radiation. 
Owing to the cavity’s high $Q$ nature, the resonance linewidth is extremely narrow ($\sim$1~pm) and thus highly sensitive to thermal drift on the scale of $\sim$10 mK, primarily governed by temperature fluctuations in both the chip and the surrounding environment. To ensure stable and repeatable measurements of the multistability, the chip is actively temperature-controlled by attaching to a thermoelectric cooler (TEC). In addition, the supporting stage and objective are enclosed within an acrylic housing to suppress convective heat exchange and air disturbances. 

As a reference, we first characterize the linear spectrum of the sample. Two distinct and comparable resonant peaks corresponding to Res A and B are observed (Fig.~3c), separated by a wavelength difference of 9 pm. A coupled-mode theory fit yields total $Q$ factors of $8.4\times10^5$ and $8.3\times10^5$ for the two modes. The balanced peak intensities and similar $Q$s indicate nearly identical radiative rates for the two resonances. Moreover, we measured the non-radiative quality factor $Q_{\rm non\text{-}rad}\sim1.2\times10^6$ from unperturbed reference samples, which reflects the random scatterings and material absorptions in the structure. It is related to the total $Q$ factor by $1/Q=1/Q_{\rm non\text{-}rad}+1/Q_{\rm rad}$, suggesting comparable rates for radiative and non-radiative losses. These observations confirm that the sample operates in the NEC regime (Fig.~2e). Such a coupling condition is strongly favorable for demonstrating optical multistability, as the enhanced field confinement and high-$Q$ resonances lead to pronounced nonlinear responses under continuous-wave excitation. Details of the fitting method and results are presented in \textcolor{black} {Supplementary Section~6}.

\section*{Experimental results of multistability}

To probe multistability, we map the reflected intensity as a function of laser wavelength and optical power (Fig.~4a). The input wavelength detuning is referenced to the midpoint between the two peaks in the linear spectrum. The resulting landscape shows the evolution from linear resonances to pronounced nonlinear behavior. The surface is constructed from combined power up- and down-scans: in gray monostable regions, the two traces overlap, while at higher power, an additional upper branch (transparent red) emerges, indicating bistability. These bistable domains originate from the cusp bifurcations of two resonances \cite{agrawal1979OpticalBistabilityNonlinear}, visible as ridge-like features that fold into bistable regions with increasing power. Transition points on the upper branch are traced by black curves, outlining cusp-shaped “horn” regions of bistability, which broaden and ultimately overlap. At this stage, the lower branch of the left resonance and the upper branch of the right resonance merge into an untraversed intermediate branch, giving rise to a multistable regime with three coexisting stable states. The hysteresis loops and transition dynamics are further confirmed from complementary parametric views.

We perform spectral cross-cutting to trace the physical origin of multistability by following the mode evolution (Fig.~4b). At a low input power of $P=\qty{2.5}{\micro\watt}$ (left panel), the spectrum exhibits a linear response, with identical results from both scan directions. As power increases, the resonant peaks shift toward longer wavelengths (middle and right panels),  as a hallmark of thermo-optical nonlinearity \cite{nishida2023OpticalBistabilityNanosilicon,carmon2004DynamicalThermalBehavior, gao2022ProbingMaterialAbsorption}. At $P=\qty{100}{\micro\watt}$ (middle panel), bistable hysteresis emerges for both resonances (highlighted red in Fig.~4a). Theoretical analysis of the underlying dynamics (see \textcolor{black}{Supplementary Section~2}) predicts the transition pathways (arrows) between branches during up- and down-scans of the input power, which are in good agreement with the experiment.

Multistability emerges when the input power exceeds \qty{240}{\micro\watt} (right panel), where the upper bistable branch of the left resonance intersects with the right one, forming a multistable regime with three coexisting branches (yellow area). Although not experimentally traversed, the theory predicts an intermediate branch (gray curve) that connects the upper and lower branches. Three stable states correspond to the cold, warm, and hot states illustrated in Fig.~2a. The measured intensities are slightly lower than theoretical predictions, especially near resonance peaks, likely due to additional nonlinear processes such as two-photon absorption and free-carrier absorption \cite{iadanza2020ModelThermoopticNonlinear,dinu2003ThirdorderNonlinearitiesSilicon,uesugi2006InvestigationOpticalNonlinearities}. Overall, the measurements are consistent with theory and reveal a multistability threshold below \qty{240}{\micro\watt}. 

We further investigate power hysteresis to directly probe equilibrium states and transition dynamics (Fig.~4c). This measurement serves as an experimental analogue of the theory in Fig.~2b, where the radiation intensity reflects the resonant energy inside the cavity and evolves in close correspondence with the temperature. Each hysteresis loop is obtained by up- and down-scanning the laser power at a fixed wavelength, corresponding to a cross-section of the input-power landscape in Fig.~4a. The measured loops (solid curves) closely match theoretical predictions (dashed curves) (see \textcolor{black}{Supplementary Section~2}), confirming the transition dynamics and validating the multistable behavior.

At a small detuning of $2$ pm (top panel in Fig.~4c), the cross-section intersects only the left resonance’s bistable branch. Off resonance, the output initially grows linearly with input power. As the power increases, the resonance shifts toward the working wavelength, giving rise to two branches corresponding to off- and on-resonance states. Further increase in power pushes the resonance past the working wavelength, restoring linearity and completing a classic bistable loop. The cross-section enters the multistable region with larger detuning, producing three-stage hysteresis loops. At $9$ pm detuning (middle panel), starting from low-power off-resonance state, increasing power first triggers the bistability of right resonance, and then the left resonance. The overlap of the two bistable regions gives rise to three coexisting stable states, corresponding to the cold, warm, and hot states in Fig.~2a. This multistability is directly confirmed by parametrically traversing through the intermediate branch, indicated by the green arrow. At $12$ pm detuning (bottom panel), the multistable region further expands, demonstrating robustness against power fluctuations. State transitions in both bistable and multistable loops follow the mechanism in Fig.~2b: during up-scans, heating shifts the resonance and triggers a rapid jump to a higher-temperature state; during down-scans, cooling reverses the shift and restores the system to a lower-temperature state once the threshold is crossed (see \textcolor{black}{Supplementary Section~2} for details).

The demonstrated multistability provides a versatile platform for optical functionalities. As a proof of concept, we realize an optical random-access memory (RAM) through time-domain switching (Fig.~5).  Memory operation is first demonstrated by modulating input power (Fig.~5a). The system is biased in the multistable regime (middle and bottom panels,  Fig.~4c), and control pulses drive transitions between states. The three intensity levels --- blue, orange, and red --- correspond to the cold, warm, and hot states, with transitions illustrated in the inset. Once a control pulse exceeds the switching threshold, the system rapidly jumps to the target state, which remains stable until the next pulse. All three states are clearly resolved and robust against noise, confirming reproducible memory behavior. Switching can also be achieved through wavelength modulation at constant power (Fig.~5b), following similar transition rules. The observed response spikes mark the resonance traversal during state changes. While the experiments are limited to dynamics of laser stabilization on the order of seconds, the intrinsic thermal response characteristic time is predicted to be $\sim \qty{3}{\micro\second}$ (see \textcolor{black}{Supplementary Section~7} for details).

\section*{Discussion}

Our findings reveal a general strategy for realizing robust optical multistability by confining multiple high-$Q$ resonant states within a single photonic system, where strong inter-resonance correlations give rise to rich nonlinear dynamics. Robustness is ensured by inherent, rather than accidental, degeneracy --- attainable through symmetry protection. Here, we harness twofold degeneracy to demonstrate optical tristability, while higher-order symmetries may naturally extend this concept toward multistability with an even greater number of accessible states. The ultra-high-$Q$ resonances not only promote light-matter interactions but also provide a versatile platform to explore diverse nonlinear mechanisms. Beyond thermo-optic effects, Kerr and other fast nonlinearities promise ultrafast, low-power operation \cite{nozaki_ultralow-power_2012,skripka_intrinsic_2025,boyd2019NonlinearOptics}.

We expect that our nonlinear near-exceptional microcavity can serve as a fundamental building block for scalable photonic computing. In particular, cascading multiple layers of such cavity arrays could enable a diffractive deep neural network \cite{lin2018all,yan2019fourier}, where interlayer communication is naturally mediated through out-of-plane radiation. This cavity architecture offers three key advantages. First, its radiative resonance allows for flexible optical interconnection between layers without the need for in-plane routing. Second, the high-$Q$ confinement significantly enhances optical nonlinearity beyond that of bulk metasurfaces, enabling each cavity to function as an efficient nonlinear activation core \cite{dubey2022activation}. Third, the realization of multistability provides access to multi-bit states that can emulate neuron-like spiking connections \cite{feldmann2019all,cheng2017chip}. Collectively, these features outline a promising pathway toward fully photonic deep-learning architectures and may bridge the gap between integrated nanophotonics and neuromorphic information processing.

\section*{Conclusion}
In this work, we demonstrate a compact and robust realization of optical multistability by controlling the radiative coupling between two-fold degenerate cavity modes operating in an NEC regime, characterized by a small wavelength splitting and balanced quality factors. Efficient excitation of resonances with $Q$s up to $8\times10^5$ is achieved, enabling the observation of clear hysteresis loops and transition dynamics governed by thermo-optical nonlinearity at an incident power below \qty{240}{\micro\watt}. Furthermore, we demonstrate a proof-of-concept optical random-access memory behavior based on the multistable states, validating the feasibility and reliability of this approach, paving the way for scalable, reconfigurable photonic memories and neuromorphic information processors.

\clearpage
\begin{figure}[htbp] 
 \centering 
 \includegraphics[width=16cm]{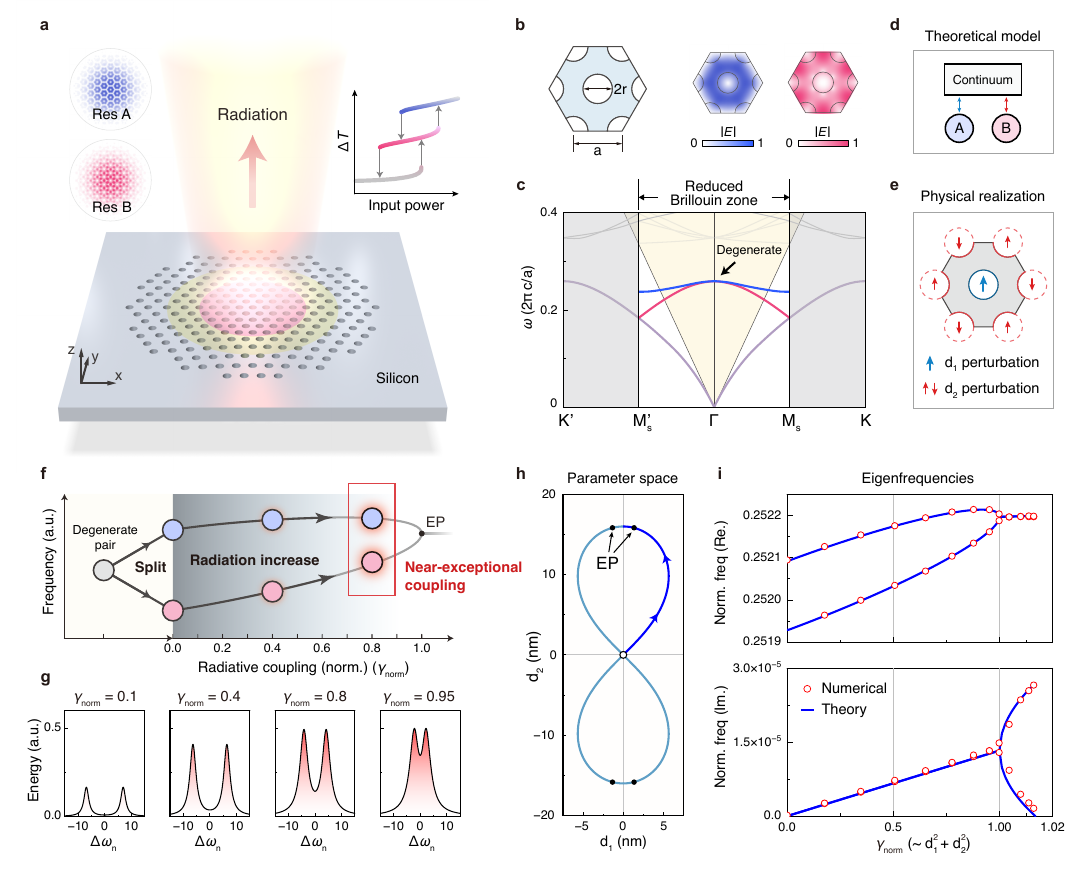}
\caption{\textbf{Photonic crystal microcavity with high-$Q$ resonances at near-exceptional coupling.}}
\label{Fig1} 
\end{figure}
\clearpage
(a) Cavity schematic highlighting core (red), transition (yellow), and cladding regions. Insets: nearly degenerate field profiles (left) and thermal multistability (right).
(b) Unperturbed PhC supercell with degenerate-mode field profiles.
(c) Primitive unit cell (light purple) and supercell (blue/red) band diagrams; $K$-point modes fold into a degenerate pair at $\Gamma$. Yellow shading marks the light cone.
(d,e) Coupled-mode model and structural implementation for tuning radiative coupling; blue and red arrows in (e) indicate hole shifts activating coupling for Res A and Res B,respectively.
(f) Eigenfrequency evolution toward the exceptional point, with radiative-coupling strength indicated by glow.
(g) Resonant spectra for varying radiative-coupling rates.
(h,i) Simulated cavity evolution: perturbation trajectory (h) and eigenfrequency traces showing EP formation (i).

\clearpage
\begin{figure}[htbp] 
\centering 
\includegraphics[width=16cm]{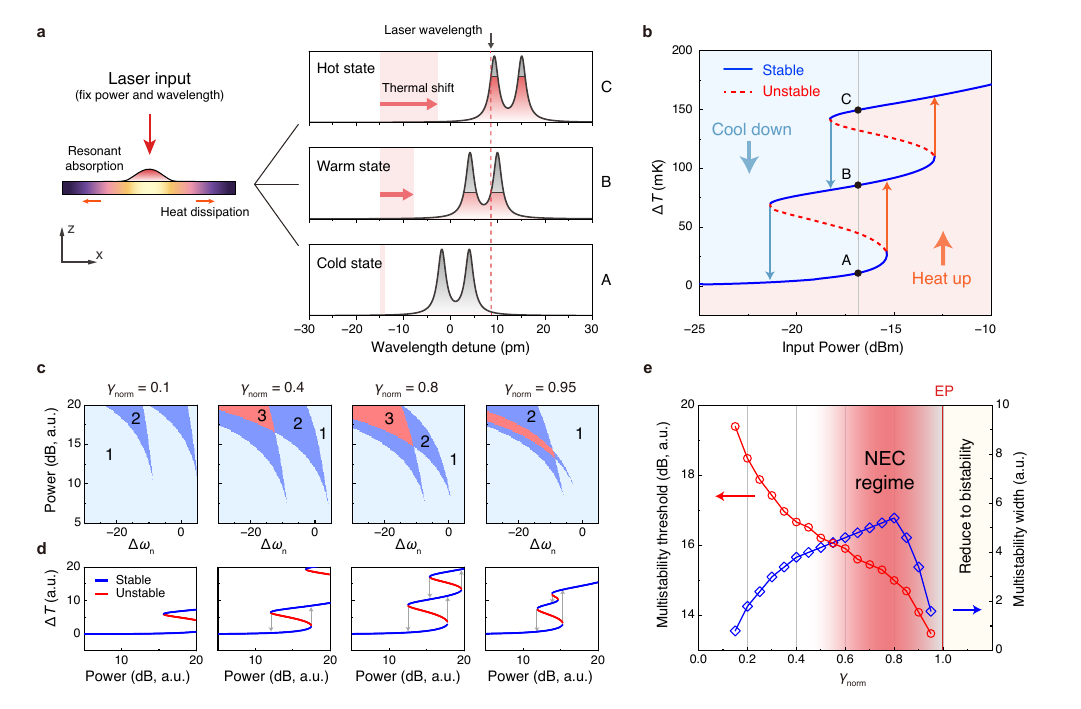}
\captionsetup{width=\textwidth}
\caption{\textbf{Thermo-optical nonlinearity under near-exceptional coupling. }}

\label{Fig2} 
\end{figure}

\clearpage
(a) Steady-state thermal response at fixed input power and wavelength (red dashed line). States A-C denote cold, warm, and hot states. Red arrows: thermal shifts; shaded regions: resonant intensity.
(b) Temperature equilibrium curves with stable (blue) and unstable (red dashed) branches. Shaded regions mark heating (red) and cooling (blue); arrows on the curve indicate transition paths.
(c) Phase diagrams at different normalized radiative coupling $\gamma_\mathrm{norm}$, showing mono-  (light blue), bi- (blue), and tri-stable (red) regions. The horizontal axis is the normalized frequency detuning $\Delta\omega_n=\Delta \omega/\gamma_\mathrm{non-rad}$.
(d) Power hysteresis at different $\gamma_\mathrm{norm}$; the first three panels use $\Delta\omega_n=-13$, while $\gamma_\mathrm{norm}=0.95$ uses $\Delta\omega_n=-10$ for clarity.
(e) Multistability threshold power (red) and width (blue) vs. $\gamma_\mathrm{norm}$. Red shading marks the near-exceptional coupling (NEC) regime. Threshold: minimum power for multistability; width: largest power range of multistability at fixed frequency under limited power (20 dB, a.u. in (c)).

\clearpage
\begin{figure}[htbp] 
\centering 
\includegraphics[width=16cm]{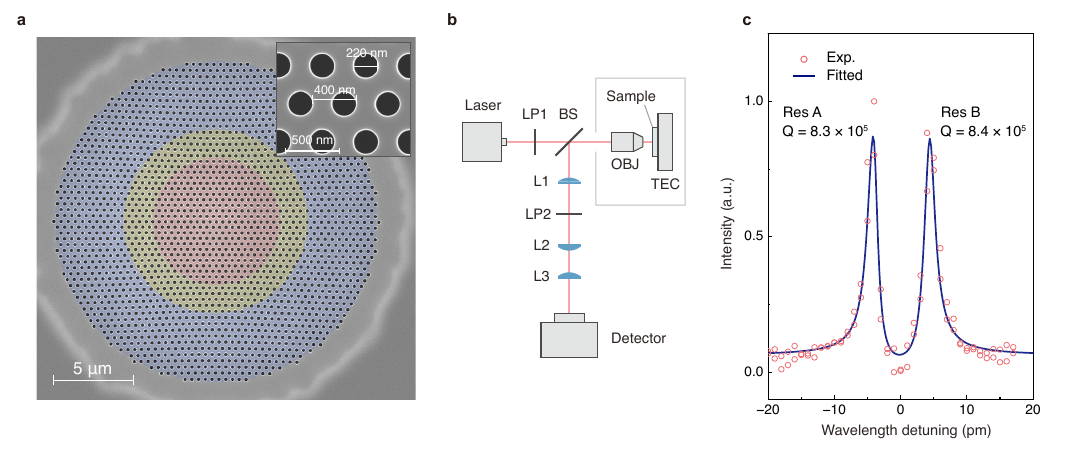}
\captionsetup{width=\textwidth}
\caption{\textbf{Realization and characterization of the optical cavity.} (a) A top-view SEM image of the cavity. The red and yellow shades denote the core and the gradient stretching region inside the blue shaded band gap region. Inset: A zoomed-in SEM image of the core region. (b) Optical setup for sample characterization. OBJ, objective; TEC, thermoelectric cooler; LP, linear polarizer; BS, beam splitter; L, lens. (c) Measured linear radiation spectrum showing two slightly split resonances with $Q$s up to $8\times10^5$,  fitted by the theoretical model.}
\label{Fig3} 
\end{figure}

\clearpage
\begin{figure}[htbp] 
\centering 
\includegraphics[width=16cm]{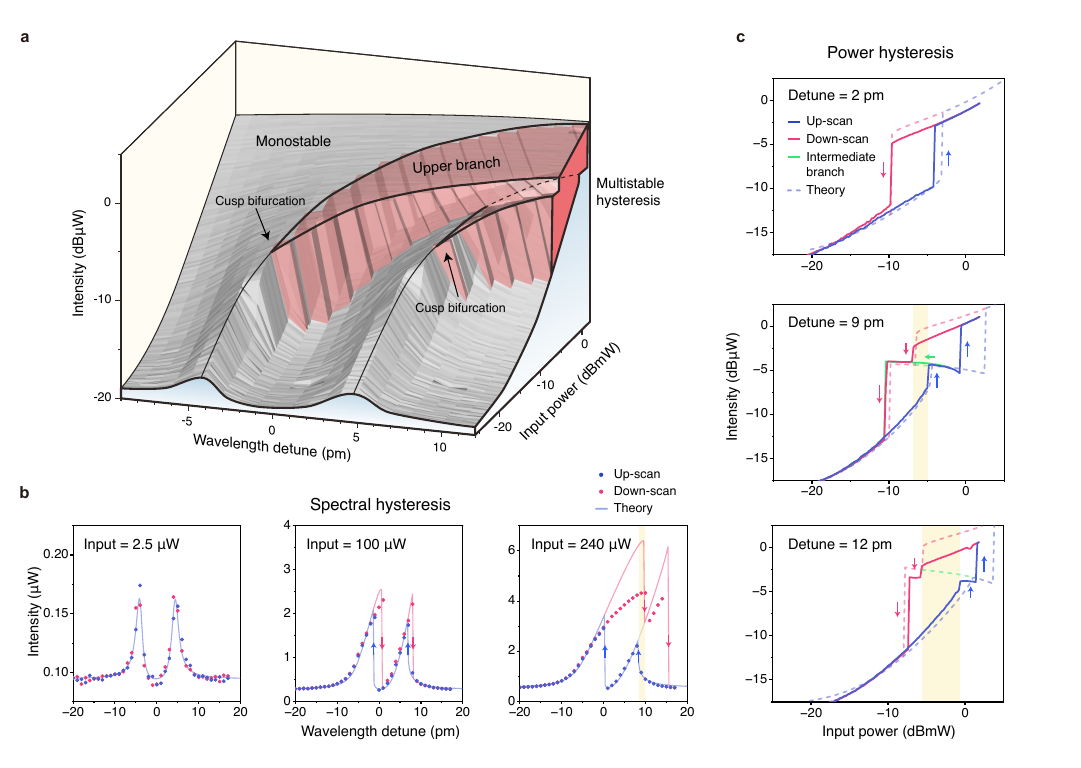}
\captionsetup{width=\textwidth}
\caption{\textbf{Experimental observation of optical multistability.}}
\label{Fig4} 
\end{figure}

\clearpage
(a) Experimental intensity (z-axis) as a function of laser wavelength (x-axis) and power (y-axis). Gray and red represent monostable and hysteretic regions; the measured upper branch of hysteresis is shown as a transparent red surface over the lower branch; and the black curves mark the cusp bifurcations.
(b) Spectral hysteresis, extracted from cross-sections along the wavelength axis at input powers of \qty{2.5}{\micro\watt} (left), \qty{100}{\micro\watt} (middle) and \qty{240}{\micro\watt} (right). Blue and red circles represent the lower and upper branches measured from power up- and down-scans, respectively; arrows mark the measured transition pathways. Solid lines denote theoretical predictions.
(c) Power hysteresis, obtained from cross-sections along the power axis at wavelength detunings of 2 pm (top), 9 pm (middle) and 12 pm (bottom). Solid lines show experimental data, with blue, red and green corresponding to up-scan, down-scan and intermediate branch, respectively; arrows indicate the scanning directions. Dashed lines denote theoretical predictions.

\clearpage
\begin{figure}[htbp] 
\centering 
\includegraphics[width=9cm]{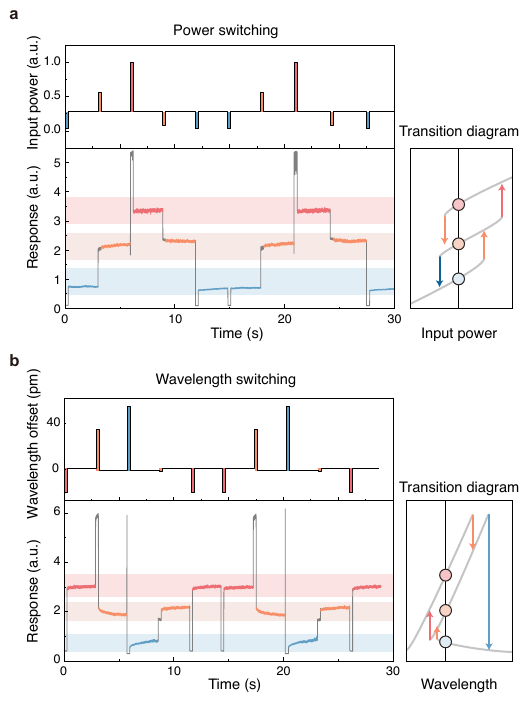}
\captionsetup{width=\textwidth}
\caption{\textbf{Demonstration of time domain switching.} 
Input signals and corresponding intensity responses under modulation of (a) input power and (b) wavelength, showing temporal switching behaviors across three states, marked by shaded regions as cold (blue), warm (orange), and hot (red) states. Insets show the transition pathways among the three states.
}
\label{Fig5} 
\end{figure}

\clearpage
\section*{Acknowledgments}
We thank Mihail Petrov, Sergey Markarov, Artem Polushkin and Fan Zhang from ITMO University for discussion. We also thank Ye Chen, Zixuan Zhang and Peishen Li from Peking University for assistance in the experiment. 
\textcolor{black}{This work was partly supported by the National Key Research and Development Program of China (2022YFA1404804) and the National Natural Science Foundation of China (62575003, 62135001, 62325501).}
\section*{Author contributions}
C.P. and Z.L. conceived the idea.
Z.L., F.W., X.Y. and C.P. performed the theoretical study and simulation. Z.L., F.W., Y.N. and Y.Z. conducted the experiments and analyzed the data. Z.L., F.W. and C.P. wrote the manuscript with input from all authors. C.P. and F.W. supervised the research. All authors discussed the results.
 
\section*{Competing interests} The authors declare no competing interests.





\clearpage
\section*{Methods}

\section*{\underline{Lattice stretching in the cavity}}
The PhC cavity consists of three layers, formed by isotropically stretching the hole positions in a 2D hexagonal lattice along the radial direction. The origin is set at a lattice point to maintain the rotational symmetry. While the angular coordinates of all air holes remain fixed, their radial positions $\rho$ are transformed into $\rho'$ according to a piecewise continuous mapping function $\phi(\rho)$, where three segments correspond to the three layers:
\[
\rho^{\prime}=\phi(\rho)=\begin{cases}\rho & \rho\le \rho_{1} \\ g(\rho-\rho_1)^2+\rho & \rho_{1}< \rho\le \rho_2 \\ s(\rho-\rho_2)+\left[g(\rho_2-\rho_1)^2+\rho_2\right] & \rho_{2}<\rho \end{cases}
\]
where $g=\frac{s-1}{2(\rho_2-\rho_1)}$; $\rho_1$ and $\rho_2$ are integer multiples of the original lattice constant, defining the core and transition region boundaries. The parameter $s$ represents the lattice scaling factor in the outer cladding region, thereby forming a quasi-periodic photonic bandgap; $s=1.125$ is used in our design. The function $\phi(\rho)$ and its first derivative are both continuous, ensuring a smooth lattice evolution across the cavity.

\section*{\underline{Numerical simulations}}
All numerical simulations are performed using the finite-element method (FEM) implemented in COMSOL Multiphysics. The cavity resonances shown in Fig.~1a and Fig.~1i are obtained by solving the eigenvalue problem in a three-dimensional cavity model with perfectly matched layers (PMLs) applied to all outer boundaries. Specifically, the results in Fig.~1i are calculated using a cavity model with a core diameter of four lattice periods, and transition and cladding regions of four and five periods, respectively, to reduce memory consumption. A fine mesh is required to capture the delicate modal evolution near the exceptional point (EP), which significantly increases computational cost. To alleviate computational burden, the central hole is enlarged by 5~nm in radius, thereby increasing the initial frequency splitting between the two resonances and effectively scaling the spectral and parameter-space features. The field distributions in Fig.~1b and the band structures in Fig.~1c are computed using a 3D supercell model with Floquet periodic boundary conditions on the sidewalls and PMLs along the vertical (top and bottom) directions.

\section*{\underline{Sample fabrication}}
The membrane PhC cavity is fabricated on a commercial silicon-on-insulator (SOI) wafer, with 220 nm top silicon layer and \qty{2}{\micro\meter} oxide layer. The cavity incorporates a three-layer heterostructure, with a core region, a photonic bandgap region, and a gradient stretching area for an adiabatic transition between the two. The whole cavity consists of 40 spatial periods in diameter, with a total size of \qty{20}{\micro\meter}. The lattice points are stretched in the radial direction to form a heterostructure cavity (see Section 1 of the Methods). 
The core has a diameter of 12 periods, and the transition and cladding regions are 4 periods and 10 periods thick, respectively. This design provides abundantly thick cladding for confinement, so the overall footprint can be further minimized in future works. Two types of perturbations are scanned in the cavities to achieve NEC: shifting the central hole ($d_1$, Fig.~1e) and the other two sets of holes ($d_2$, Fig.~1e) in each supercell. We make a two-dimensional parameter matrix to scan two perturbations simultaneously. The PhC cavities are patterned on the SOI chip by electron beam lithography (Elionix ELS-F125), followed by inductively coupled plasma etching with CHF$_3$ and SF$_6$ (Oxford Cobra180). After stripping the electron resist, the chip is immersed in $49\%$ hydrofluoric acid at room temperature for one minute to remove the underneath insulator layer and form a membrane cavity. 

\section*{\underline{Measurement and data processing}}
The incident light source is a tunable laser (Santec TSL-550) operating between 1500 and 1630 nm. We focus the incident light onto the sample using a $\times50$ objective lens (IOPAMI137150X-NIR, Mitutoyo). To reduce background noise, we employ a cross-polarized detection scheme. The excitation and detection paths are filtered using linear polarizers LP1 and LP2, aligned vertically and horizontally, respectively. The sample is oriented at $45^\circ$ relative to the horizontal, enabling simultaneous excitation and detection of linearly polarized resonances. The chip is placed on a thermally conductive stage connected to a thermoelectric cooler (TEC). The sample stage and objective are enclosed in an acrylic box to minimize convective heat exchange and air disturbances. To characterize the linear spectrum in Fig.~3c, the incident wavelength is continuously scanned while the photodetector (PD) records the signal. The spectrum is then achieved during the wavelength scanning. To characterize the multistability in Fig.~4a, the incident wavelength is scanned in a point-by-point manner. At each wavelength, the power is up-scanned and then down-scanned for recording. To perform the time-switching experiment in Fig.~5, the signal from PD is recorded continuously while the laser wavelength (or power) is modulated under computer control. The input optical power is measured as the net beam power incident on the sample. The output power is defined as the optical power entering the objective lens from the sample, corrected for losses in the subsequent light path.

\section*{Data availability}
All the data supporting the findings of this study are available in the main paper and its Supplementary Information. Additional information can be obtained from the corresponding authors upon request. 

\end{document}